# Experimental confirmation of the redundancy of the axiomatic principles of statistical physics

Vladimir V. Savukov

In the process of analyzing the axiomatic principles underlying statistical physics, when modeling the most probable stationary macrostates of non-ergodic closed systems, a forecast was obtained about a possible limitation purview of the main postulate, known as the "principle of equiprobability for each realizable of microstate". It is assumed that if such a system is artificially brought into a thermodynamically equilibrium state, in particular, by filling it with thermal Planck radiation of the appropriate temperature, then the subsequent evolution of the state of this system can cause the appearance of stable polarization anisotropy of said radiation. The paper presents the successful results of a direct verification of aforesaid prognosis on a real physical installation. Important regularities characteristic of the applied mathematical model are noted.

# РЕФЕРАТ

В процессе анализа аксиоматических принципов, лежащих в основе статистической физики, при моделировании наиболее вероятных стационарных макросостояний неэргодических замкнутых систем был получен прогноз о возможной ограниченности сферы действия основного постулата, известного как «принцип равновероятности для каждого реализуемого микросостояния». Предполагается, что если такую систему искусственно привести в термодинамически равновесное состояние, в частности, заполнив ее тепловым планковским излучением соответствующей температуры, то последующая эволюция состояния этой системы может вызвать появление устойчивой поляризационной анизотропии указанного излучения. В статье представлены успешные результаты прямой проверки вышеизложенного прогноза на реальной физической установке. Отмечены важные закономерности, характерные для применяемой математической модели.

**Ключевые слова:**  статистическая физика, аксиоматика, эргодичность, негэнтропия, угол Брюстера, поляризация.

## Experimental confirmation of the redundancy of the axiomatic principles of statistical physics

Vladimir V. Savukov

# ABSTRACT

In the process of analyzing the axiomatic principles underlying statistical physics, when modeling the most probable stationary macrostates of non-ergodic closed systems, a forecast was obtained about a possible limitation purview of the main postulate, known as the "principle of equiprobability for each realizable of microstate". It is assumed that if such a system is artificially brought into a thermodynamically equilibrium state, in particular, by filling it with thermal Planck radiation of the appropriate temperature, then the subsequent evolution of the state of this system can cause the appearance of stable polarization anisotropy of said radiation. The paper presents the successful results of a direct verification of aforesaid prognosis on a real physical installation. Important regularities characteristic of the applied mathematical model are noted.

# СОДЕРЖАНИЕ

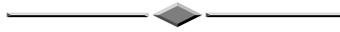

# Введение

Настоящая статья публикуется в рамках поискового проекта, посвящённого анализу границ применимости аксиоматических принципов статистической физики [1-3, 5-10, 12]. Существующий аппарат статистической физики равновесных систем базируется на гипотезе о равновероятности всех микросостояний, доступных рассматриваемой замкнутой системе [1]. При этом имеется в виду следующее:

1. Наиболее вероятное стационарное состояние замкнутой (изолированной от внешней среды) физической системы называется равновесным состоянием. Равновесное состояние является макроскопическим. Оно представляет собой совокупность всех доступных системе микросостояний, т. е. таких конкретных состояний, каждое из которых может быть осуществлено при заданном уровне энергии.

2. В каждый фиксированный момент времени равновесное состояние реализуется через одно из составляющих его микросостояний. При этом система с одинаковой вероятностью может быть обнаружена в любом из микросостояний, образующих её равновесное макросостояние.

Пункт 2 данной аксиоматики позволяет декларировать то, как именно должно выглядеть равновесное состояние замкнутой системы, а также определяет направленность стохастических процессов во времени. Последнее означает, что, например, взаимодействие фотонов теплового излучения внутри такой системы с любым находящимся в ней оптическим элементом не способно изменить макроскопические параметры этого излучения, если они уже соответствуют описанному определению равновесного состояния.

Ранее [1-3] было выдвинуто предположение о существовании неэргодических[1] квантовых систем, поведение которых лежит вне "зоны ответственности" статистической физики. Это обусловлено тем, что фазовая траектория каждой квантовой частицы[2] не является непрерывной на уровне подпространства импульсов. Способность квантовых частиц «исчезать» и «появляться» в различных частях доступного им фазового пространства открывает возможность существования в этом пространстве источников и стоков фазовых траекторий, имеющих не одинаковую плотность в одних и тех же локальных участках фазового объёма. Возникающая в результате устойчивая во времени ненулевая дивергенция потока фазовых траекторий в конкретных частях фазового пространства может сделать данную систему неэргодичной, а её свойства — не сообразными с аксиоматикой статистической физики.

Сказанное дало основания допустить, что при определённых условиях диффузный фотонный газ[3] может менять исходное изотропное макросостояние на анизотроп-ное, которое в этих условиях будет более вероятным. Иначе говоря, микросостояния фотонов термодинамически равновесного излучения, изначально равномерно заполняющие некоторый объём фазового пространства, могут быть в нём перераспределены вследствие "расщепления" фазовых траекторий единичных фотонов на множественные когерентные каналы рассеяния, например, на дифракционные порядки – после взаимодействия излучения с решёткой. Если решётка является составной частью замкнутой физической системы, то больцмановская энтропия такой системы способна уменьшаться с течением времени. Данный парадокс преодолевается путём определения энтропии через формулу Шеннона. Это позволяет формулировать понятие энтропии, как меры вероятности макросостояния замкнутой системы, не прибегая к постулату о равновероятности её микросостояний, используемому при дефиниции энтропии Больцмана [12].

---

[1] Свойство эргодичности предполагает достоверность микроканонической гипотезы статистической физики о тождественности результатов усреднения по времени и фазового усреднения при вычислении значений макроскопических параметров системы.

[2] Применительно к квантовым частицам понятие фазовой траектории может быть сохранено путём его переопределения на основании теоремы Эренфеста [1].

[3] Под диффузным фотонным газом здесь понимается неполяризованное некогерентное электромагнитное излучение, для отдельных фотонов которого равновероятна любая возможная угловая ориентация в трёхмерном геометрическом пространстве их волновых **k**-векторов.

Далее излагается результат компьютерного моделирования наиболее вероятных стационарных макросостояний неэргодических замкнутых систем, в которых термодинамически равновесное планковское излучение спонтанно приобретает анизотропную поляризацию, а также подводится итог натурных экспериментов по успешному выявлению данного эффекта на реальной физической установке[1].

### Система на основе фазовой дифракционной решётки и её недостатки

Предполагалось, что вышеупомянутый эффект может использоваться при пассивной локализации объектов, находящихся в термодинамическом равновесии с окружающей средой (например, в скрытых системах безопасности и т. п.). Для этого данные объекты должны быть "маркированы" дифракционными решётками, поверхность которых становится видимой при наблюдении их через тепловизор с поляризационным фильтром [1]. Однако описанное техническое решение имеет ряд существенных недостатков. Главная проблема заключается в том, что анизотропия поляризационных параметров диффузного фотонного газа, возникающая после его взаимодействия с дифракционной решёткой, хорошо проявляется, если данный газ монохромный. Но состояние термодинамического равновесия характеризуется тепловым излучением с планковским распределением частот. При этом дифракционные порядки рассеяния фотонов, принадлежащих различным участкам спектра, в очень значительной степени (около 95-98%) компенсируют друг друга на уровне суммарной энергетической яркости. Ранее даже предполагалось [1], что такая взаимная компенсация может достигать 100%, и в этом случае носит обязательный характер, принципиально не позволяющий тепловизорам с матрицами болометрического типа зафиксировать прогнозируемый эффект[2].

На рис. 1 приведён ряд графических изображений, иллюстрирующих данное обстоятельство. Каждый график построен в полярной системе координат так, что его центр соответствует нулевому значению угла отражения при внешнем обзоре поверхности дифракционной решётки. Величина угла отражения пропорциональна полярному радиусу, и на периферии графика значение этого угла приближается к 90°. Азимутальный угол наблюдения поверхности решётки определяется полярным углом.

---

[1] Ранее было подтверждено наличие подобного эффекта для монохромного случая [2].

[2] Данное предположение явилось причиной того, что в первой экспериментальной установке [2] объектом исследования стал предварительно стохастизированный монохромный фотонный газ, а не реальное тепловое излучение с планковским спектром.

Исходное световое поле представляет собой диффузное излучение с общим числом фотонов в статистическом испытании $N = 285\,749\,842$. Индикатриса **1**а описывает угловое распределение яркости монохромного (длина волны $\lambda = 10$ мкм) диффузного светового поля, отражаемого от идеально проводящей фазовой линейной решётки (шаг $d = 8.200$ мкм, полная глубина синусоидального профиля микрорельефа $h = 3.116$ мкм, штрихи микрорельефа ориентированы вертикально). Этот график автоматически масштабируется так, чтобы максимальным образом выявлять все имеющиеся контрасты плотности рассеянного светового потока. На изображении индикатрисы **1**а присутствуют лишь бессистемные проявления флуктуаций, в соответствии с законом Ламберта не образующие каких-либо устойчивых макроскопических градиентов [1].

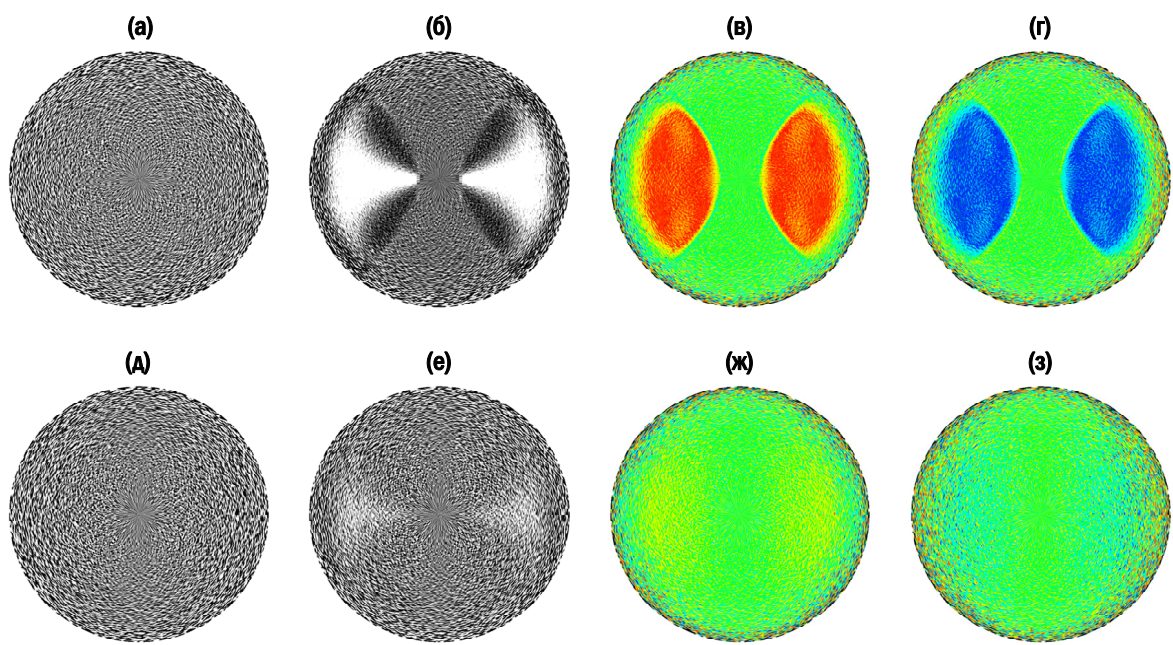

Рис. 1. Поляризационные характеристики изначально диффузного светового поля, рассеиваемого фазовой дифракционной решёткой с целью пассивной локации маркированного объекта:
 – Верхний ряд (1а-1г) содержит информацию для монохромного светового поля;
 – Нижний ряд (1д-1з) содержит информацию для светового поля с планковским спектром.

На рис. **1**б приведено изображение расчётной плотности вероятности угла поляризации α, определяемого как арктангенс отношения амплитуд взаимно ортогональных компонент электрического вектора в произвольной системе координат [11]. Этот график содержит сильно выраженные градиенты, вызываемые дифракцией монохромного излучения на отражательной решётке.

На рис. **1**в и **1**г представлены изображения, соответственно, $S$- и $P$-индикатрис, которые, согласно компьютерному прогнозу, можно наблюдать на экране тепловизора, снабжённого поляризационным фильтром (анализатором), – как результат дифракции монохромного диффузного излучения на решётке.

Для сравнения на рис. **1** приведён второй ряд графических изображений (**1**д, **1**е, **1**ж и **1**з), содержащих информацию, аналогичную размещённой в первом ряду (см. рис. **1**а, **1**б, **1**в и **1**г), но соответствующую случаю не монохромного, а планковского спектра изотропного фотонного газа, полностью отвечающего определению термоди-намически равновесного излучения с температурой 290°K. Содержание графиков **1**ж и **1**з свидетельствует о том, что рассмотренная методология эксперимента практически не пригодна для работы с реальным излучением, обладающем протяжённым (планковским) спектром. Использование же на входе в тепловизор наряду с анализатором ещё и частотного узкополосного фильтра (для выделения монохромной компоненты), скорее всего, приведёт к фатальному падению величины анализируемого сигнала, исходный уровень которого и так невелик при типичной температуре окружающей среды ~300°K.

Из других недостатков использования дифракционных решёток для поляризации изначально термодинамически равновесного излучения следует упомянуть ещё два:

– Характер распознаваемого тепловизором сигнала сильно зависит от того, под каким ракурсом видна поверхность решётки. Есть такие сочетания углов отражения и азимутальных углов, которые образуют, своего рода, "мёртвые зоны", исключающие обнаружение поляризационных градиентов даже в среде монохромного излучения.

– Особые требования к геометрии микрорельефа дифракционной решётки могут существенно увеличивать её стоимость, поскольку наряду с шагом и глубиной этого микрорельефа регламентируется строго синусоидальная форма его профиля [2]. Кроме того, данный микрорельеф весьма уязвим перед любым внешним воздействием и может быть легко повреждён. Использование же каких-либо защитных покрытий на поверхности решётки способно существенно снизить или даже полностью исключить проявление рассматриваемого поляризационного эффекта.

## Неэргодическая система на базе плоского диэлектрического зеркала

В ходе имитационного моделирования наиболее вероятных макросостояний замкнутых физических систем обнаружилось, что дифракционная поляризация на регулярных структурах не является единственным механизмом, способствующим появлению негэнтропийных процессов в таких системах. Подобные возможности выявлены и у конструкций, в которых в качестве оптических элементов вместо решёток используются диэлектрические зеркала. В этом случае разрыв фазовых траекторий частиц, необходимый для появления у системы неэргодических свойств, происходит в процессе преодоления фотонами границы между внутренним объёмом зеркала и внешней средой.

Рассмотрим простейший вариант замкнутой системы с оптическим элементом в виде плоского диэлектрического зеркала. Компьютерная модель такой системы, изначально находящейся в состоянии термодинамического равновесия, прогнозирует спонтанное возникновение анизотропной поляризации, заметной при наблюдении поверхности зеркала под углом отражения, равном углу Брюстера. Указанная анизотропия заключается в возникновении диспропорций между $S$- и $P$-компонентами фиксируемого излучения[1], что может быть выявлено при его фильтрации анализатором.

Ожидаемый эффект должен быть тем сильнее, чем больше значение коэффициента преломления у материала диэлектрического зеркала. Наиболее подходящие для этой цели германий Ge и селенид цинка ZnSe имеют коэффициенты преломления, которые в пределах их окон внутреннего пропускания мало зависят от частоты излучения. Таким образом, различным частотам планковского спектра будут соответствовать примерно одни и те же величины угла Брюстера[1], т. е., в отличие от поляризации на дифракционной решётке, здесь самокомпенсация проявлений анизотропии будет выражена слабо.

---

[1] Суммарная энергетическая яркость $S$- и $P$-компонент при этом остаётся неизменной.

Для максимально полного соответствия свойств исследуемой имитационной модели прогнозируемым характеристикам реальной физической системы, было решено использовать особый фактор, стохастизирующий параметры фотонного газа в толще диэлектрического зеркала (см. описание германиевого оптического Ge-окна к рис. 5.1).

На рис. 2 дана принципиальная схема физической установки, предназначенной для выявления прогнозируемого эффекта (её конкретную реализацию см. на рис. 5).

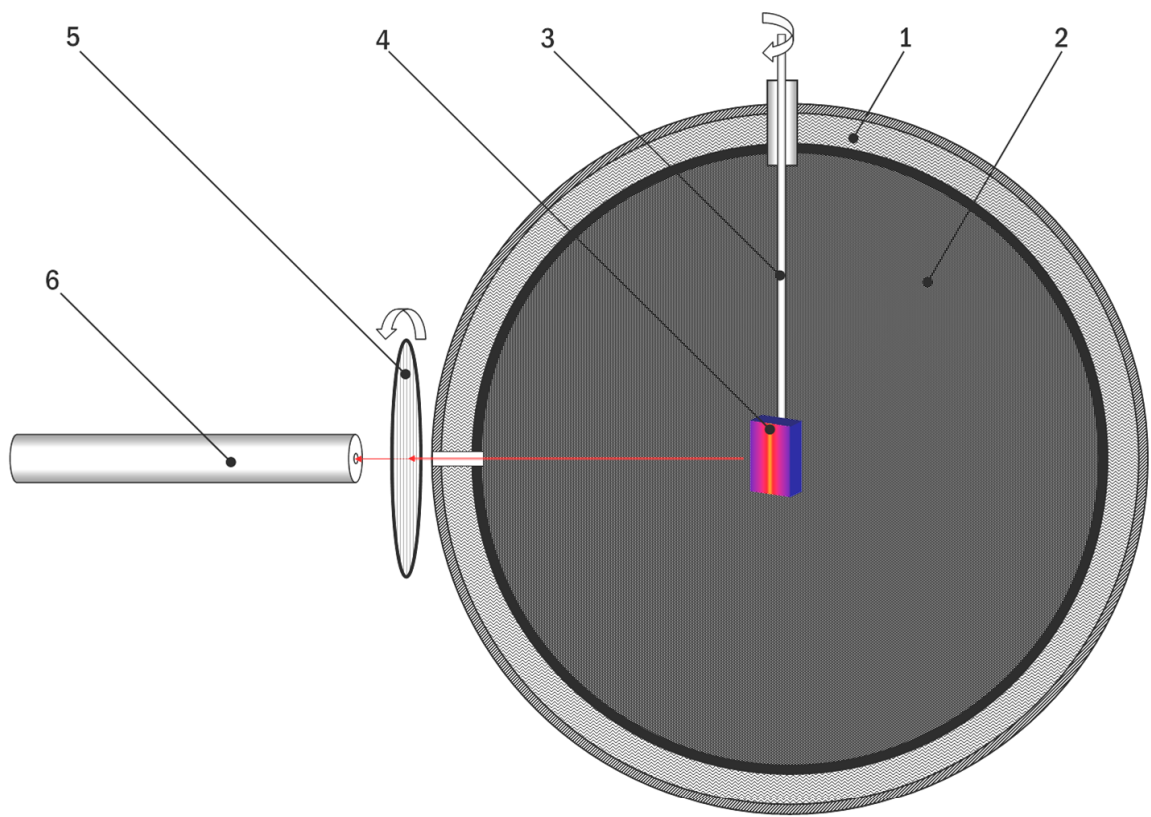

**Рис. 2.** Схема экспериментальной установки на основе плоского диэлектрического зеркала

### Перечень элементов установки:

1. Термостатированный фотометрический полый шар, ограничивающий квазизамкнутую систему[2]. Оболочка шара, содержащая толстый слой теплоизоляции, минимизирует воздействие внешних тепловых потоков на размещённые внутри него объекты.

2. Внутренняя поверхность шара 1, покрытая поглощающим материалом со свойствами, близкими к свойствам абсолютно чёрного тела. Это, например, может быть

---

[1] Угол Брюстера $\theta_{Br}$ зависит от отношения коэффициентов преломления материала диэлектрического зеркала $n$ и окружающей его среды $n_0$: $\theta_{Br} = \arctan(n/n_0)$.

[2] Здесь имеется ввиду то, что квазизамкнутый характер физической системы допускает получение внешним наблюдателем информации о её внутреннем состоянии.

Vantablack™ – специальная субстанция из углеродных нанотрубок, характеризуемая полным интегральным коэффициентом отражения ~ 0.045% [4].

3. Цилиндрический стержень, способный вращаться вокруг своей оси. Служит для размещения прикреплённого к нему диэлектрического зеркала 4 в центре шара 1. Изменение угла, под которым оптическая ось входной апертуры регистрирующего прибора 6 ориентирована к плоскости зеркала 4, выполняется поворотом стержня 3.

4. Основной оптический элемент, представляющий собой плоское диэлектрическое зеркало из материала с высоким коэффициентом преломления и окном внутреннего пропускания шириной, как минимум, 8-14 мкм (наиболее подходит германий Ge). Зеркало не должно иметь просветляющих и иных покрытий на рабочей поверхности!

5. Анализатор инфракрасной области излучения в диапазоне 8-14 мкм. Должен иметь возможность поворота вокруг своей оптической оси. Компоненты излучения, отфильтрованные для регистрирующего прибора 6, определяются сочетанием угла пропускания этого анализатора с текущей ориентацией диэлектрического зеркала 4.

6. Регистрирующий прибор, например, тепловизор с рабочим диапазоном $\lambda \approx$ 8-14 мкм или низкотемпературный (~300°К) радиационный пирометр полного излучения. Вариант с тепловизором обладает большей наглядностью фиксируемой информации. Он лучше подходит для получения результатов качественного характера (см. рис. 5).

Данная экспериментальная установка способна выявить избыток энергетической яркости $S$-компоненты или соответствующий провал яркости $P$-компоненты[1] в составе радиационного потока, фиксируемого прибором 6 в процессе вращения зеркала 4. Эти вариации, согласно расчёту, будут достигать величины ±4.0% для зеркала из селенида цинка ZnSe (угол Брюстера ≈ 67.4°) и ±5.6% – для зеркала из германия Ge (угол Брюстера ≈ 76.0°). Отклонения столь значительного масштаба соответствуют таким макроскопическим градиентам температур, которые могут быть надёжно зарегистрированы, например, бытовым тепловизором или пирометром с умеренной чувствительностью.

### Система с диэлектрическим зеркалом сферической формы

Использование диэлектрического зеркала сферической формы позволяет одновременно наблюдать на участках его поверхности с различным азимутом как избыток, так и провал яркости у поляризационных $S$- и $P$-компонент, содержащихся в составе теплового излучения, идущего со стороны указанных участков под углом Брюстера. В качестве регистрирующего прибора следует применить тепловизор, снабжённый поляризационным фильтром. Прогнозируемый эффект в данном случае должен быть одинаково хорошо заметен при любом угловом ракурсе между тепловизором и зеркалом.

На рис. 3 приведены сравнительные результаты моделирования взаимодействия сферического германиевого зеркала как с монохромным, так и с планковским диффузным излучением, причём порядок и смысловое содержание полярных графиков здесь соответствуют тем, которые были ранее представлены на рис. 1.

Индикатриса 3а описывает угловое распределение яркости монохромного (длина волны $\lambda$ = 10 мкм) диффузного светового поля, фиксируемого для всех наблюдаемых ракурсов сферического зеркала. На изображении индикатрисы видны только проявления флуктуаций, не создающие каких-либо устойчивых макроскопических градиентов.

---

[1] Эти соотношения прогнозируются относительно равных пропорций поляризационных компонент, что имеет место при наличии термодинамического равновесия в системе.

На рис. **3**б дан график расчётной плотности вероятности угла поляризации α. Это изображение содержит выраженные макроградиенты, обусловленные диспропорцией между поляризационными *S*- и *P*-компонентами в потоке излучения, идущего со стороны диэлектрического зеркала под углом Брюстера[1] к его поверхности.

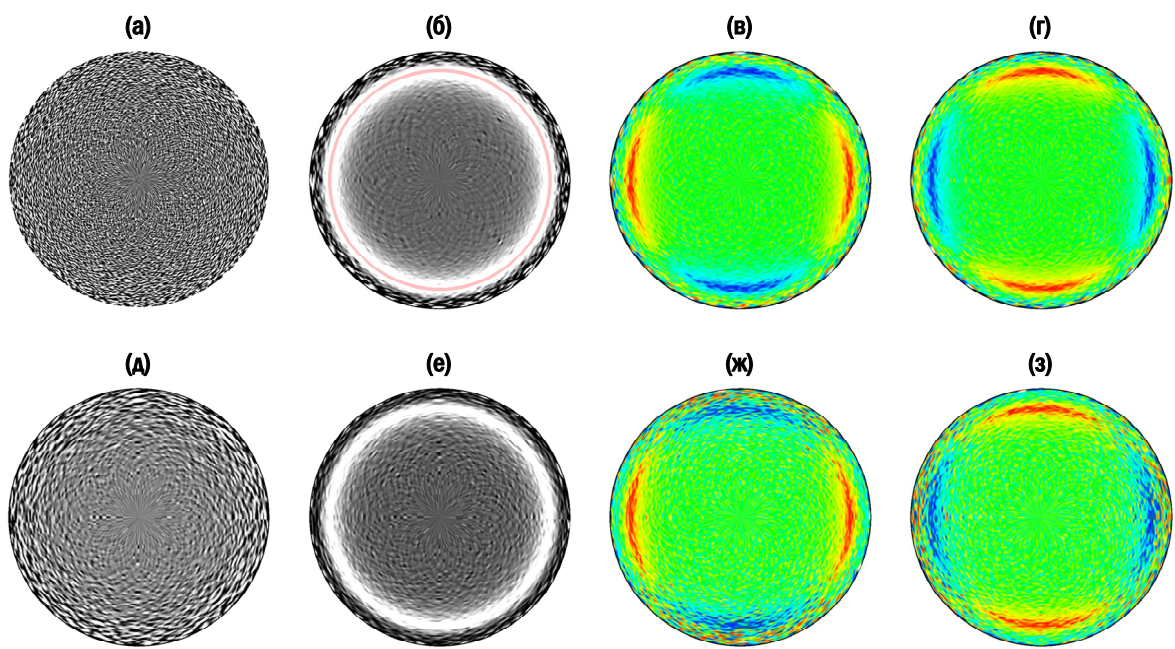

**Рис. 3. Поляризационные характеристики, аналогичные приведённым на рис. 1, но для случая использования сферического диэлектрического зеркала вместо дифракционной решётки.**

На рис. **3**в и **3**г представлены изображения, соответственно, *S*- и *P*-индикатрис, которые, согласно компьютерному прогнозу, можно наблюдать на экране снабжённого анализатором тепловизора как результат поляризации монохромного диффузного излучения с помощью диэлектрического зеркала.

Для сравнения на рис. **3** приведён второй ряд графических изображений (**3**д, **3**е, **3**ж и **3**з), содержащих информацию, аналогичную показанной в первом ряду (см. рис. **3**а, **3**б, **3**в и **3**г), но соответствующую случаю не монохромного, а планковского спектра изотропного фотонного газа, отвечающего определению термодинамически равновесного излучения с температурой 290°К. Содержание графиков **3**ж и **3**з свидетельствует о том, что методология эксперимента с использованием диэлектрического зеркала вполне пригодна для работы с реальным излучением, обладающем протяжённым спектром.

Эффективность проявления негэнтропийного эффекта может быть проиллюстрирована на примере прогнозируемых результатов применения в экспериментах диэлектрических зеркал сферической формы, выполненных из материалов с различными коэффициентами преломления "*n*". На рис. **4** представлены ожидаемые картины, которые в ранее заданных условиях (равновесное излучение, T = 290°К) можно будет наблюдать на экране тепловизора, снабжённого поляризационным фильтром.

---

[1] Углы отражения от поверхности диэлектрического зеркала, равные углу Брюстера, отмечены на полярном графике рис. **3**б красной кольцевой линией.

На рис. 4а дан прогноз изображения *S*-компоненты поляризованного излучения, идущего со стороны диэлектрического зеркала, выполненного из монокристаллического германия Ge. Ниже этого изображения приведён график (рис. 4г) относительной яркости[1] той части излучения, которая наблюдается под углом отражения θ (вертикальной чертой отмечено значение угла Брюстера $θ_{Br} ≈ 75.98°$). Для сравнения на рис. 4б и 4д представлены аналогичные характеристики зеркала из поликристаллического селенида цинка ZnSe, а на рис. 4в и 4е – для зеркала из фтористого бария $BaF_2$. Ожидаемая вариация температур, фиксируемая тепловизором относительно исходного значения *T* = 290°K, составит для зеркала из германия ±5.59°K ($n ≈ 4.00$), для зеркала из селенида цинка ±4.03°K ($n ≈ 2.41$) и для зеркала из фтористого бария ±0.03°K ($n ≈ 1.40$).

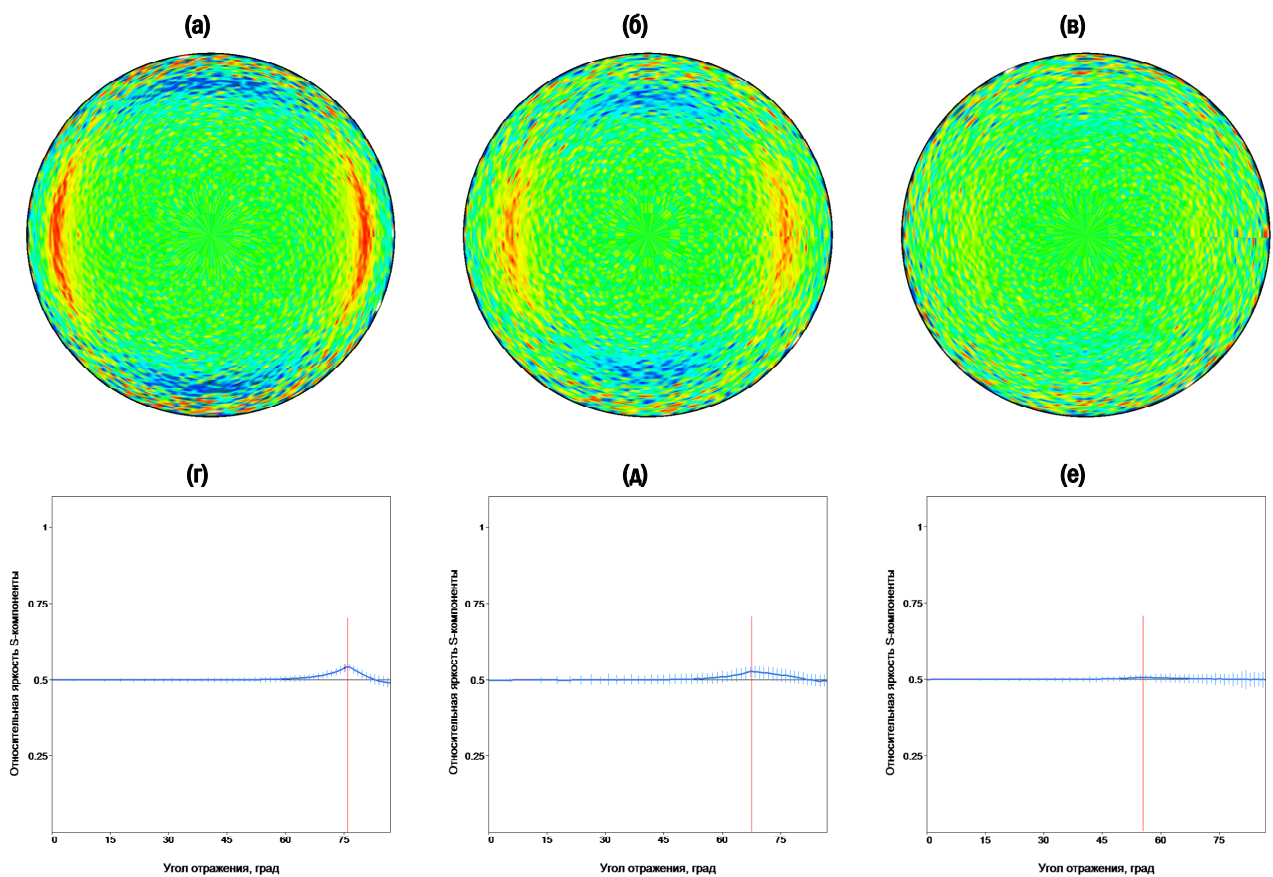

**Рис. 4.** Прогнозируемые изображения *S*-индикатрис для зеркал, выполненных из Ge ("а"), ZnSe ("б") и $BaF_2$ ("в"), а также графики радиальных сечений этих индикатрис в горизонтальной плоскости.

Если, например, дисперсию на графике 4г считать мерой достоверности значения математического ожидания энергетической яркости для Ge, то её девиация может быть флуктуационным проявлением равновесного состояния с вероятностью менее $10^{-4}$.

### Прямая экспериментальная проверка существования прогнозируемого эффекта

Проверка самого факта существования ожидаемого эффекта была выполнена по модифицированной схеме, а именно: в качестве основного оптического элемента вме-сто образца сферической формы было использовано так называемое "окно" – плоская пластина из монокристаллического германия оптического качества. На рис. 5 даны фотоснимки частей собираемой экспериментальной установки, представляющей собой квазизамкнутую физическую систему, исходное состояние которой было искусственно «приготовлено» близким к состоянию термодинамического равновесия (ΔT ≈ ±0.2°K).

---

[1] В состоянии термодинамического равновесия относительная доля энергетической яркости каждой из поляризационных компонент – тождественно равна ½.

**1. Германиевое "окно"**   **2. Азимутальная карта**   **3. Николь - анализатор**   **4. Тепловизор SATIR D300**

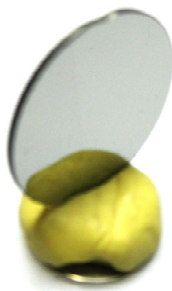 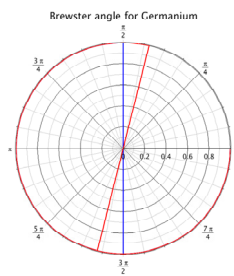 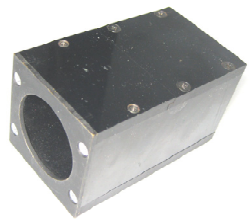 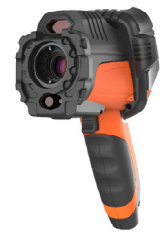

**5. Положение Ge-окна (1), размещённого на карте (2) под углом Брюстера к входной апертуре анализатора (3)**

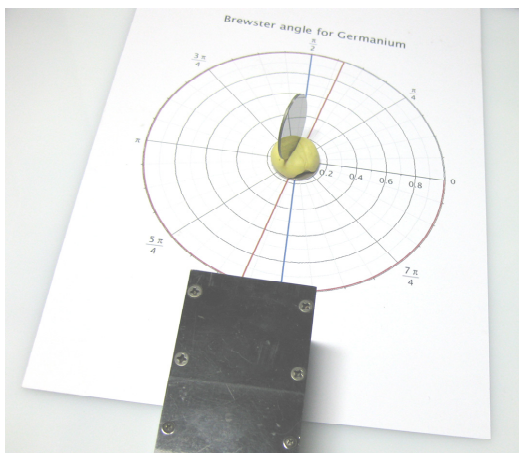 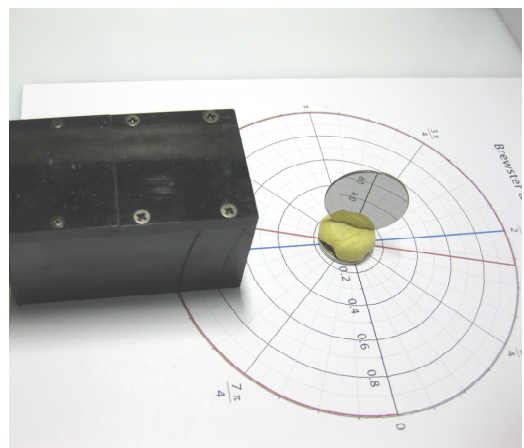

**Рис. 5.** Компоненты установки, предназначенной для обнаружения прогнозируемого эффекта в реальной физической системе с умерено однородным температурным полем (ΔT ≈ ±0.2°K)

## Перечень комплектующих элементов установки:

1. Германиевое "окно" – плоский диск из оптически чистого германия. Его рабочая сторона полирована и не имеет каких-либо просветляющих или защитных покрытий. Обратная сторона Ge-окна специально матирована для создания на ней хаотического микрорельефа. Это сделано с целью стохастизации фотонного газа в толще данного образца, что приближает параметры физического эксперимента к расчётной модели.

2. Азимутальная карта – графическое изображение индикатрисы, служащее для задания угла отражения, под которым анализатор (см. п. 3) "видит" поверхность Ge-окна.

3. Николь - анализатор – прибор, выполненный в технологии Николь - призмы. Имеет такую особенность, как выделение только одной наблюдаемой поляризационной компоненты из всего потока излучения, что может стать очень значимым фактором!

4. Тепловизор "SATIR D300" – прибор фирмы SATIR™ Europe (Ireland) LTD.

5. Снимки, поясняющие взаимное угловое расположение элементов 1, 2 и 3 установки.

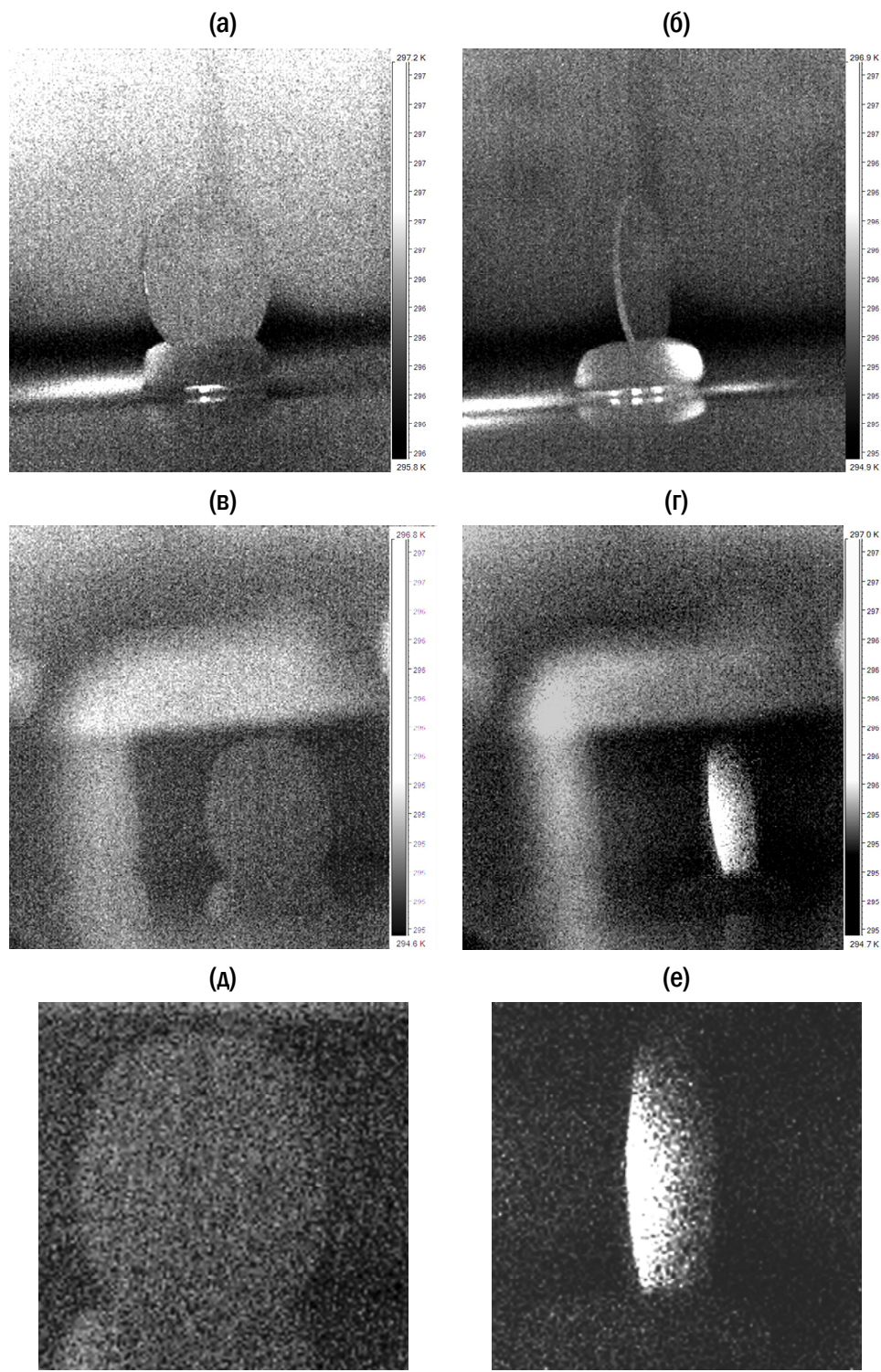

Рис. 6. Спонтанная поляризация собственного излучения квазизамкнутой физической системы:
– а: Ge-окно в открытом кофре (без анализатора), ракурс наблюдения его поверхности 45°
– б: Ge-окно в открытом кофре (без анализатора), его поверхность видна под углом Брюстера
– в: передняя панель анализатора, Ge-окно стоит под углом 45° к его входной апертуре
– г: передняя панель анализатора, угол Брюстера между его входной апертурой и Ge-окном
– д: снимок крупного плана выходной апертуры анализатора, угловой ракурс Ge-окна 45°
– е: крупный план выходной апертуры анализатора, Ge-окно видно под углом Брюстера ≈ 76°

### Пояснения к результатам эксперимента, представленным на рис. 6

На рис. 6 приведены два основных результата выполнения эксперимента в максимально изолированной (практически замкнутой) физической системе: например, для защиты установки от внешних излучений применялся специальный кофр и т. д. Исход каждого из этих экспериментов определялся выбранным значением угла отражения, под которым входной канал анализатора "видел" рабочую поверхность Ge-окна (см. рис. 5.5):

– угол отражения 45° (см. рис. 6а, 6в и 6д) – не приводил к какому либо отклонению от изотропности поляризационных свойств планковского излучения, характеризующего исходное состояние термодинамического равновесия физической системы;
– угол отражения, равный углу Брюстера (для Ge ≈ 76°, см. рис. 6б, 6г и 6е) – вызывал устойчивое нарушение изотропности поляризационных параметров в итоговом, наиболее вероятном (но не термодинамически равновесном) состоянии данной системы.

Осуществление эксперимента в таком избыточно строгом варианте (см. рис. 6), когда система максимально возможным образом изолирована от окружающей среды, является необходимым шагом для проверки корректности получаемых результатов. Однако после успешного выполнения этой проверки становится допустимым обоснованное смягчение условий проведения работ с реальными оптическими элементами.

Например, отказ от использования любого освещения во время съёмок приводил к проблемам при автофокусировке на Ge-окно, когда оно пребывает в термодинамическом равновесии со средой, что визуально проявляется в форме бесструктурного шума (см. рис. 6в и 6д). Проблема была решена путём применения слабого служебного освещения, направляемого со стороны тепловизора перпендикулярно передней панели анализатора. При этом автофокусировка осуществлялась на фронтально освещаемые (и потому хорошо видимые на фоне стохастических флуктуаций) конструктивные элементы во внутреннем канале анализатора. В то же время используемые в экспериментах углы отражения от Ge-окна (45° или 76°) практически исключали увеличение яркости этого окна при освещении, направляемом со стороны тепловизора. Германиевое окно является эффективным диэлектрическим зеркалом. В указанных ракурсах оно отражает такой фронтально ориентированный свет либо поперёк оптической оси анализатора (если угол отражения был 45°, см. рис. 6а), либо вообще в противоположном от тепловизора направлении (если угол отражения был равен углу Брюстера: 76°, см. рис. 5.5 и рис. 6б). Это позволило получить чёткие снимки искомых артефактов[1] (см. рис. 7).

### Пояснения к результатам эксперимента, показанным на рис. 7

На рис. 7 изображены четыре типичных результата эксперимента. Исход каждого из них определялся выбранным сочетанием двух следующих угловых характеристик:
– угла отражения, под которым анализатор "видит" рабочую поверхность Ge-окна,
– угла поворота поляризационной плоскости пропускания Николь-анализатора.

---

[1] Здесь под *артефактом* понимается некий процесс, который в текущих условиях принято считать невозможным (по аксиоматическим причинам) или чрезвычайно маловероятным (флуктуации). Визуальное проявление такого артефакта представляет собой инфракрасное изображение поверхности германиевого окна, имеющего форму диска (см. рис. 5.1) и наблюдаемого почти параллельно плоскости данного диска (т. е. под углом Брюстера 76°). Прогнозируется, что при этом ракурсе энергетическая яркость поляризационной *S*-компоненты на 5% превысит яркость такой же компоненты, излучаемой физической системой в состоянии её термодинамического равновесия.

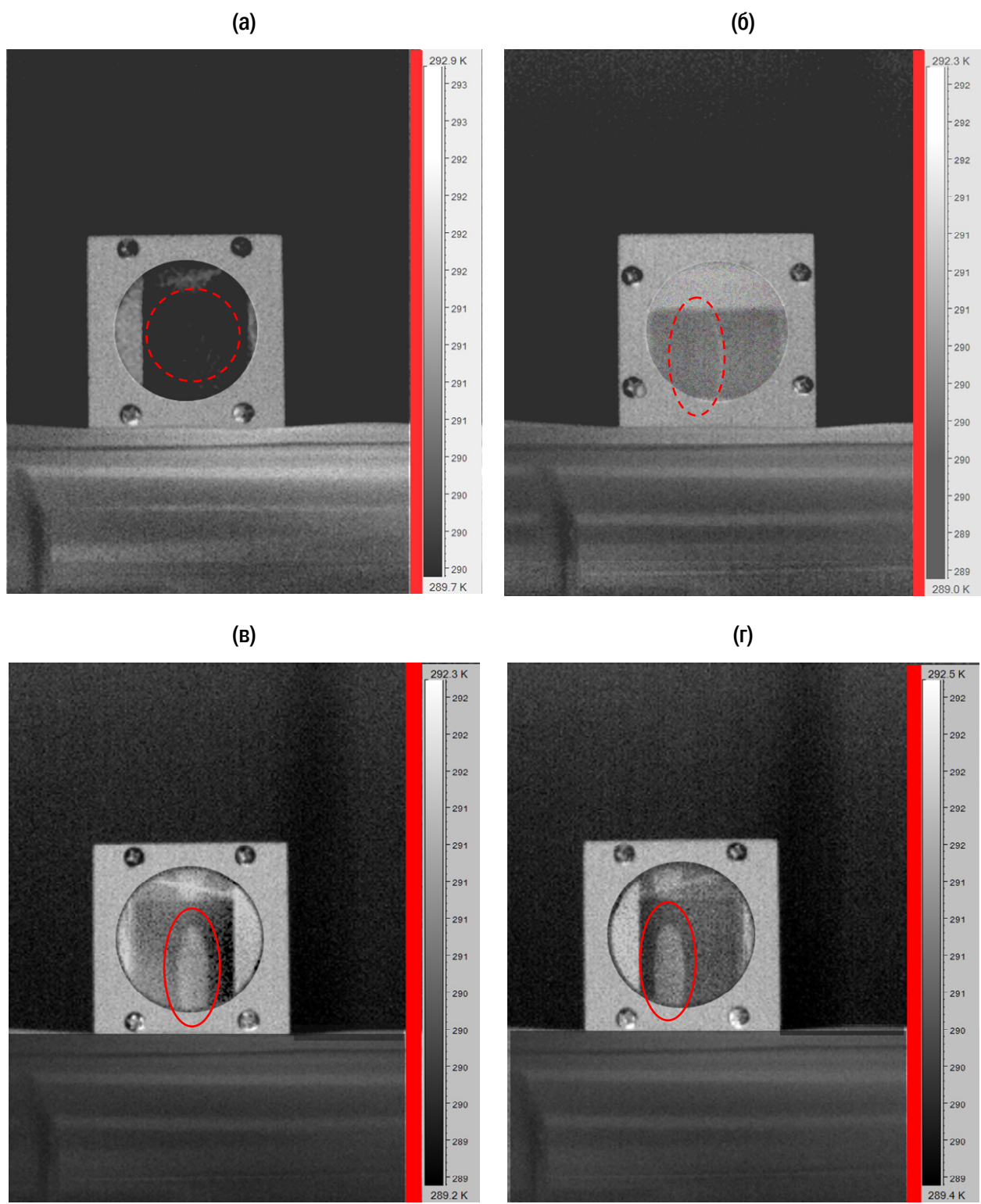

Рис. 7. Спонтанная поляризация термодинамически квазиравновесного планковского излучения, взаимодействующего с пластиной германиевого окна в реальной физической системе:
  – а: угол поворота оси анализатора: 180°, поверхность Ge-окна видна под ракурсом 45°
  – б: угол поворота оси анализатора: 090°, поверхность Ge-окна видна под ракурсом 76°
  – в: угол поворота оси анализатора: 000°, поверхность Ge-окна видна под ракурсом 76°
  – г: угол поворота оси анализатора: 180°, поверхность Ge-окна видна под ракурсом 76°

- а. Угол поворота анализатора[1] на 180° формально позволяет зафиксировать анизотропную поляризацию в составе излучения, идущего от оптического элемента. Однако сама данная поляризация здесь отсутствует ввиду того, что угол отражения исследуемого потока от Ge-окна назначен отличным от угла Брюстера (45° вместо 76°).

- б. Угол поворота анализатора на 90° не даёт возможности увидеть наличие анизотропной поляризации в составе излучения, исходящего от оптического элемента. При этом сама упомянутая анизотропия тут реально существует, поскольку угол отражения тепловой радиации от Ge-окна принят равным углу Брюстера. Однако проявление данной анизотропии для такого угла поворота Николь-анализатора связано с поляризационной компонентой, которая принудительно гасится в Николе. На рис. **7**б красной пунктирной линией обведено предполагаемое место на апертуре, где визуально проявился бы артефакт, как если бы применяемый Николь-анализатор не поглощал вторую поляризационную компоненту проходящего через него излучения[2].

- в. Угол отражения исследуемого потока от Ge-окна принят равным углу Брюстера. Угол поворота анализатора на 0° по отношению к плоскости пропускания даёт возможность наблюдать анизотропную поляризацию в составе излучения, идущего от оптического элемента (этот артефакт обведён на рис. **7**в сплошной красной линией).

- г. Угол отражения внешней тепловой радиации от Ge-окна здесь также принят равным углу Брюстера. Поэтому угол поворота анализатора на 180° к плоскости пропускания тоже обеспечивает возможность увидеть картину анизотропной поляризации (соответствующий артефакт обведён на рис. **7**г сплошной красной линией).

Таким образом, результаты всех физических экспериментов находятся в согласии с прогнозами, ранее полученными при имитационном моделировании данных работ.

### Обнаруженные закономерности параметров фазового пространства

Все проявления анизотропии параметров фотонного газа, ранее обнаруженные при анализе наиболее вероятных стационарных состояний замкнутых систем, обусловлены исключительно только вариациями угла поляризации α в "упруго" формируемом световом поле[3] (см., например, рис. **1**б и **3**б). Однако, это не означает, что изменения угла α могут носить произвольный характер. Результаты моделирования итоговых макросостояний замкнутых систем, получаемые после завершения всех переходных процессов, непременно характеризуются следующим обязательным свойством[1]:

$$\int_0^{\pi/2} \cos(\alpha)^2 g(\alpha) d\alpha \equiv \frac{1}{2} \tag{1}$$

где $g(\alpha)$ - это плотность вероятности угла поляризации, т. е. нормированная к единице неотрицательная функция; для состояния термодинамического равновесия она имеет единственно возможный (в соответствии с аксиоматикой статистической физики) вид:

$$g(\alpha) = \sin(2\alpha) \geq 0, \quad \int_0^{\pi/2} g(\alpha) d\alpha = 1, \quad \forall \alpha \in [0, \pi/2] \tag{2}$$

---

[1] Следует отметить, что угол поворота анализатора является здесь менее однозначной характеристикой, чем угол отражения от Ge-окна. Если угол отражения будет хотя бы на 2-3 градуса отличаться от брюстеровского значения, то искомый эффект уже никак не проявит себя. Что же касается конкретных значений углов пропускания и запирания для анализатора, то они сильно зависят как от конструкции самого анализатора, так и от характера поляризационной анизотропии исследуемого потока излучения.

[2] Подобная амплитудная невзаимность допускает возможность направленного переноса энергии излучения между разными частями термодинамически равновесной системы.

[3] Имеется в виду взаимодействие светового поля с оптическими элементами системы, осуществляемое без необратимых потерь на поглощение. Их учёт потребовал бы применения аксиоматики неравновесных процессов (гипотезы молекулярного хаоса [**12**]). Законы Кирхгоффа (баланс испускаемого и поглощаемого излучения), Ламберта (угловые распределения испускаемого и отражаемого излучений) и Планка (частотный спектр равновесного излучения) при этом полностью сохраняют свою корректность.

Разумеется, при такой дефиниции плотности вероятности невозможно появление каких-либо градиентов у количественного соотношения *S*- и *P*-компонент в системе. Попробуем теперь ввести модифицированный вариант плотности вероятности $G(\alpha)$, который бы отвечал тождеству (1) и нормировке (2), но при этом позволял варьировать угол поляризации α максимально свободным образом:

$$G(\alpha) = g(\alpha) + R(\alpha) \geq 0, \quad \forall \alpha \in [0, \pi/2] \tag{3}$$

Разложим в ряд Фурье функцию $R(\alpha)$, добавленную в состав $G(\alpha)$:

$$R(\alpha) = \frac{a_0}{2} + \sum_{k=1}^{\infty} \left( b_k \sin(k\alpha) + a_m \cos(m\alpha) \right) \tag{4}$$

Тогда ограничения на гармоники, разрешённые для функции $R(\alpha)$, выглядят так:

$$\int_0^{\pi/2} G(\alpha) d\alpha = \int_0^{\pi/2} \sin(2\alpha) + R(\alpha) d\alpha \equiv 1 \tag{5}$$

$$\int_0^{\pi/2} \cos^2(\alpha) R(\alpha) d\alpha \equiv 0, \quad \forall a_m \tag{6}$$

Окончательный вид функции $G(\alpha)$ с учётом условий (5) и (6):

$$G(\alpha) = \sin(2\alpha) + \sum_{m=4,6\ldots}^{\infty} a_m \cos(m\alpha) \geq 0, \quad \forall \alpha \in [0, \pi/2] \tag{7}$$

На рис. 8 приведены графики для четырёх случаев плотностей вероятности углов поляризации. Синими линиями отмечены графики равновесных состояний $g(\alpha)$. Кружками обозначены данные имитационных компьютерных моделей. Красными линиями выполнена аппроксимация этих данных с помощью функции $G(\alpha)$, область допустимых значений (ОДЗ) которой содержит лишь около одной четверти всех без исключения гармоник ряда Фурье (4). Тем не менее, разрешённых гармоник (7) оказывается вполне достаточно для того, чтобы точно соответствовать моделируемым параметрам, учитывая даже проявления флуктуаций. Такой результат однозначно доказывает принадлежность рассчитываемых состояний к множеству, соответствующему условию (1).

---

[1] Свойство (1) не актуально, если система не замкнута или далека от стационарности.

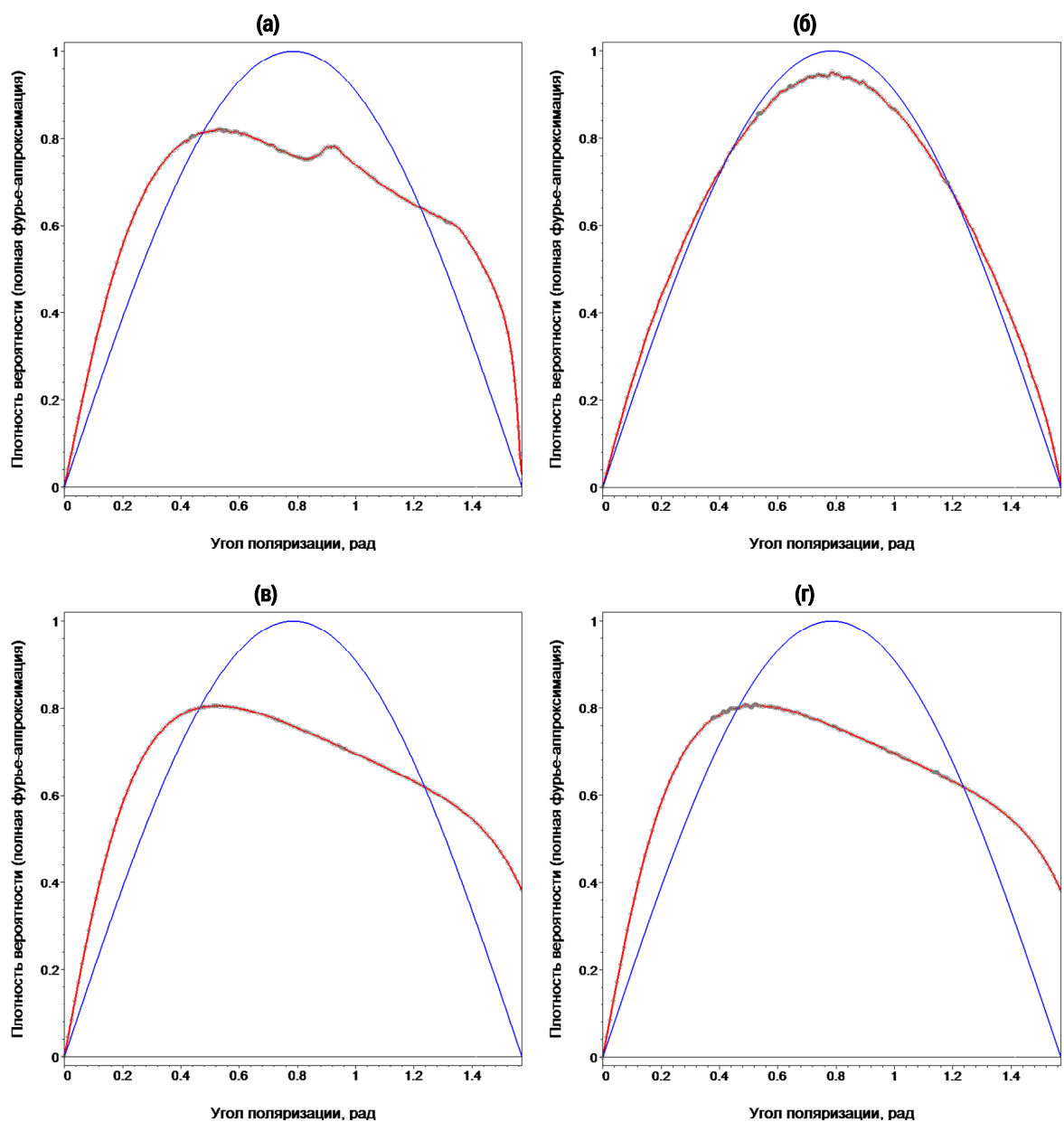

Рис. 8. Графики плотности вероятности угла поляризации $G(\alpha)$ в его области допустимых значений:
  а). $G(\alpha)$ - монохромный фотонный газ после рассеяния на отражающей фазовой решётке
  б). $G(\alpha)$ - планковский фотонный газ после рассеяния на отражающей фазовой решётке
  в). $G(\alpha)$ - монохромный фотонный газ после взаимодействия с диэлектрическим зеркалом
  г). $G(\alpha)$ - планковский фотонный газ после взаимодействия с диэлектрическим зеркалом

Полученные данные иллюстрируют ранее сделанные выводы о различной эффективности отражающих фазовых дифракционных решёток и диэлектрических зеркал для поляризации диффузного фотонного газа с планковским спектром. На рис. 8а и 8в видно, насколько существенно отличаются плотности вероятностей $G(\alpha)$, формируемые из исходных распределений $g(\alpha)$ для планковского фотонного газа, и для монохромного излучения. Анизотропная трансформация излучения с планковским спектром проявляется большей частью лишь при использовании в качестве оптического элемента диэлектрического зеркала (рис. 8г), но не регулярной структуры типа решётки (рис. 8б).

Надо особо отметить значимость свойства (**1**) для замкнутых систем с той точки зрения, что, несмотря на возможность существенной анизотропии угла поляризации в геометрическом пространстве (**7**), эта анизотропия сама по себе принципиально неспособна привести к появлению градиентов температур в замкнутой системе. Хотя, например, для осмия Os разница в коэффициентах поглощения *S*- и *P*-компонент термодинамически равновесного излучения источника типа "А" (T = 2856°К) может достигать почти двукратной величины, никакая анизотропная поляризация, если она проявляется в сочетании с условием (**1**), не способна изменить коэффициент поглощения µ. Это происходит потому, что зависимость данного коэффициента от угла поляризации выражается линейной параметрической функцией, в которой одной из переменных выступает коэффициент поглощения µ, а в качестве другой − квадрат косинуса угла α:

$$\mu(\alpha) = \mu_1 \eta(\alpha) + \mu_0, \quad \eta(\alpha) = \cos^2(\alpha) \qquad (0.1)$$

где $\mu_0$ и $\mu_1$ - константы, определяющие свойства материала, поглощающего излучение.

Таким образом, если в фазовом пространстве замкнутой системы реализуется условие (**1**), то величина коэффициента поглощения µ начинает совпадать со значением, характерным для изотропного излучения, не имеющего поляризационных градиентов. Это исключает возникновение разницы температур в такой системе, если только для получения температурного перепада не будет задействован какой-либо иной механизм.

## Выводы

В ходе работ, выполненных по этому проекту, был получен следующий результат: теоретически обосновано и экспериментально подтверждено существование неэргодических квазизамкнутых физических систем, наиболее вероятные стационарные макросостояния которых зависят от их внутренней организации. Для внешнего наблюдателя такая девиация проявляется в форме устойчивой поляризационной анизотропии теплового излучения, заполняющего данные системы в указанных макросостояниях.

Планируемые работы по данной тематике состоят в развитии полученного результата, допускающего, что термодинамическое равновесие не является единственно возможным наиболее вероятным макросостоянием для неэргодических систем. Вышеизложенное ставит под сомнение всеобъемлющий характер главного аксиоматического принципа статистической физики о равновероятности всех доступных микросостояний в замкнутой системе, а это − не исключает ревизии границ корректного применения Н-теоремы Больцмана, представляющей собой статистический аналог Второго закона термодинамики. Разумеется, оценка перспектив, вытекающих из сказанного, должна быть крайне сдержанной. Однако, до сих пор верификация исследуемой имитационной модели не выявила существенных ошибок, а прогнозируемые особые эффекты нашли своё подтверждение в положительных результатах натурных экспериментов. Поэтому продолжение поисковых работ в принятом направлении следует считать оправданным.

## Список использованной литературы

1. ***Савуков В. В.*** Анизотропная поляризация, прогнозируемая как результат дифракции излучения чёрного тела на отражающей фазовой решётке с идеальной проводимостью // Оптический журнал, 2012, том **79**, № 10, с. 7-15.
   URL: http://www.savukov.ru/opticjourn_79_10_2012_rus.pdf
   URL: http://www.savukov.ru/opticjourn_79_10_2012_eng.pdf

2. ***Савуков В. В.*** Экспериментальное подтверждение негэнтропийного характера дифракционной поляризации диффузного излучения // Оптический журнал, 2016, том **83**, № 12, с. 3-9.
   URL: http://www.savukov.ru/opticjourn_83_12_2016_rus.pdf
   URL: http://www.savukov.ru/opticjourn_83_12_2016_eng.pdf

3. ***Савуков В. В., Голубенко И. В.*** Моделирование взаимодействия произвольного светового поля с дифракционной решёткой методом Монте-Карло // Оптический журнал, 2012, том **79**, № 7, с. 10-17.
   URL: http://www.savukov.ru/opticjourn_79_07_2012_rus.pdf
   URL: http://www.savukov.ru/opticjourn_79_07_2012_eng.pdf

4. ***Yang Zu-Po, Ci Lijie, Bur James A., Lin Shawn-Yu, Ajayan Pulickel M.*** Experimental Observation of an Extremely Dark Material Made By a Low-Density Nanotube Array // Nano Letters, Vol. **8**, Issue 2, pp. 446–451. (9 January 2008).

5. ***Синай Я. Г.*** Динамические системы с упругими отражениями (эргодические свойства рассеивающих бильярдов) // Успехи математических наук, 1970, том **25**, № 2 (152), с. 141-192.

6. ***Геворгян А. А.*** Невзаимность волн в поглощающих многослойных средах // Письма в ЖТФ, том **29**, вып 19.

7. ***Курилкина С. Н.*** Оптическая невзаимность для встречных световых волн при дифракции Брэгга в гиротропных кубических кристаллах // Квантовая электроника, 1995, том **22**, № 9, с. 941-945.

8. ***Сычёв А. Н., Малютин Н. Д.*** Современные устройства, антенны и отражатели с невзаимными свойствами (обзор) // Журнал радиоэлектроники, 2020, № 11.

9. ***Веселовская Т. В., Клочан Е. Л., Ларионцев Е. Г., Парфёнов С. В., Шелаев А. Н.*** Амплитудная и фазовая невзаимности акустооптических модуляторов для встречных световых волн при дифракции Брэгга // Квантовая электроника, 1990, том **17**, № 7, с. 823-829.

10. ***Кравцов Н. В., Кравцов Н. Н.*** Невзаимные эффекты в кольцевых лазерах // Квантовая электроника, 1999, том **27**, № 2, с. 98-120.

11. ***M. Born, E. Wolf*** Principles of optics. Electromagnetic Theory of Propagation, Interference and Diffraction of Light // Pergamon Press (Fourth Edition) 1970, 859 p.

12. ***Савуков В. В.*** Уточнение аксиоматических принципов статистической физики // Деп. ВИНИТИ РАН. № 1249-В2004 от 16.07.2004.
    URL: http://www.savukov.ru/viniti_1249_b2004_full_rus.pdf